\documentclass[floatfix, twocolumn, prb]{revtex4}

\usepackage{color}
\usepackage{multirow}
\usepackage{epsfig}
\usepackage{graphicx}
\usepackage{amsmath}
\usepackage{amssymb}
\usepackage{bm}
\usepackage{dcolumn}
\usepackage{braket}

\begin{document}

\title{Topological phases of Bi(111) bilayer in an external exchange field}

\author{Hongbin~Zhang$^1$}
\email[corresp.\ author: ]{h.zhang@fz-juelich.de}
\author{Frank Freimuth$^1$}
\author{Gustav Bihlmayer$^1$}
\author{Stefan~Bl\"ugel$^1$}
\author{Yuriy~Mokrousov$^1$}
\affiliation{$^1$Peter Gr\"unberg Institut and Institute for Advanced Simulation, 
Forschungszentrum J\"ulich and JARA, D-52425 J\"ulich, Germany}

\date{today}

\begin{abstract} 
Using first principles methods, we investigate topological phase transitions
as a function of exchange field in a Bi(111) bilayer.
Evaluation of the spin Chern number for different magnitudes of the exchange field
reveals that when the time reversal symmetry is broken by a small exchange field, 
the system enters the time-reversal broken topological insulator phase, introduced by Yang {\it et al.} in Phys. Rev. Lett. 107, 066602 (2011).
After a metallic phase 
in the intermediate region,  the quantum anomalous Hall phase with non-zero 
Chern number emerges at a sufficiently large exchange field. 
We analyze the phase diagram from the viewpoint of the evolution of the 
electronic structure, edge states and transport properties, and demonstrate 
that different topological phases can be distinguished by the spin-polarization of the
edge states as well as spin or charge transverse conductivity.
\end{abstract}

\maketitle

\section{Introduction}

Topological insulators (TIs) have drawn quite intensive attention recently owing  
 to the gapless surface/edge states, topologically  protected against
perturbations which do not close the bulk gap.\cite{Hasan:2010, Qi:2011}
Similarly to graphene,\cite{Neto:2009} exciting physical properties are expected if
the TI surface/edge states are made gapped, resulting in~e.g.~half-quantized surface Hall 
effect,\cite{Qi:2008} topological magneto-electric effects,\cite{Qi:2009, Essin:2009} 
and Majorana zero states.~\cite{Fu:2008} It is suggested that such effects can be triggered
via  the proximity effect of TIs to magnetically ordered materials or superconductors, in
which case new terms which impose or break certain symmetries are introduced into the
effective Hamiltonian. 

One particularly interesting situation arises if the time-reversal symmetry ($\mathcal{T}$) 
in a TI is broken via a controlled perturbation. The definition of the $\mathbb{Z}_2$ topological 
invariant, used to classify two-dimensional (2D) insulators into ordinary insulators and TIs, is 
hinged on the presence of the time-reversal symmetry in the system.\cite{Fu:2007} It still remains 
an open question of how to generalize the $\mathbb{Z}_2$ topological invariant when $\mathcal{T}$ 
is broken. In such a case, it is suggested\cite{Yang:2011} that it is plausible to use the so-called 
spin Chern number:\cite{Prodan:2011} 
\begin{equation}
\mathcal{C}_s=\frac{1}{2}(\mathcal{C}_+\ -\ \mathcal{C}_-),
\end{equation}
where $\mathcal{C}_{+}$ and $\mathcal{C}_{-}$ are the Chern numbers for the spin "up" and "down" 
manifolds of the occupied states. Since the spin-orbit coupling (SOC) in general induces spin mixing, 
the spin Chern number is well defined only when the system exhibits a gap in the spectrum of 
$\sigma_z$ in addition to being a band insulator. In this case $\mathcal{C}_s$ proves to be 
equivalent to the $\mathbb{Z}_2$ number when time-reversal symmetry is present.\cite{Prodan:2010}
For models of TIs\cite{Prodan:2011} and graphene,\cite{Yang:2011} the spin Chern number picture 
proved to be valid, while a more detailed analysis of this concept in real materials investigated from 
first principles is still lacking.

Owing to $\mathcal{T}$ symmetry, the surface/edge states of TIs are spin degenerate, making it 
hard to manipulate and exploit their transport properties in experiments based on magnetic detection. 
While the quantum Hall effect usually requires a very large magnetic field to emerge, in this sense, the  
quantum anomalous Hall (QAH) effect in ferromagnets would be much more suited for future spintronics. 
The QAH effect, which has attracted a lot of attention recently but so far has not been verified experimentally,
occurs in 2D Chern insulators\cite{Haldane:1988} with broken $\mathcal{T}$-symmetry, exhibiting overall 
spin-polarized edge states carrying quantized  electric charge. Up to now, it is believed that TIs can be turned 
into Chern insulators via magnetic doping\cite{Liu:2008, Yu:2010} but it still remains an experimentally 
unsolved problem despite recent advances in this direction.\cite{Chang:2011} In this light, it is necessary 
to explore how tunable 2D TIs are under  $\mathcal{T}$-broken perturbations, in particular, exchange 
interactions.

Bismuth is the heaviest atom with a stable isotope and strong spin-orbit coupling
makes it an important ingredient of the newly discovered  TIs such as Bi$_2$Se$_3$\cite{Xia:2009} 
and Bi$_{1-x}$Sb$_x$.\cite{Fu:2007} Ultrathin Bi(111) films, which can be produced
experimentally on different substrates,\cite{Nagao:2004, Hirahara:2011} are also 
predicted to be 2D TIs.\cite{Murakami:2006, Wada:2011, Liu:2011} In our study we take
Bi(111) bilayer as a representative of the latter class of TIs. In this work, we investigate the 
topological phase transitions in a 2D TI -- the Bi(111) bilayer\cite{Wada:2011} $-$ in an external 
exchange field using first principles methods. 
We found that at small exchange fields the bilayer is in the time-reversal broken 
topological insulator phase, which can be characterized by non-trivial values of spin Chern numbers, 
transverse spin Hall conductivity, and spin polarization of the edge states.
Increasing the magnitude of the exchange field further drives the system first into a 
metal and then into a QAH phase. Detailed analysis of the electronic structures reveals that spin-mixing 
is important for closing the bulk gap and Chern number exchange between valence and conduction bands. 
At last, we analyze the spin-polarization of the metallic edge states in different phases, show that its sign 
can be controlled by the strength of the exchange field,  and prove that its magnitude is  large enough
to be observed in scanning tunneling microscopy experiments. 

\section{Method} 
In a Bi(111) bilayer two layers of Bi atoms form a honeycomb lattice when projected onto the plane 
of the film. The relaxed bulk in-plane lattice constant and the distance between the two layers were 
4.52~\AA~and 1.67~\AA, respectively (see also Fig.~4).  Our theoretical investigations are based on 
density functional theory.\cite{Hohenberg:1964} We apply the local density approximation~\cite{Moruzzi:1978} 
to the exchange-correlation potential and use the full-potential linearized augmented plane wave method 
(FLAPW) as implemented in the {\tt FLEUR} code.\cite{fleur} The self-consistent calculations with SOC were 
carried out with the cut-off parameter k$_{\rm max}$ of 3.8~bohr$^{-1}$ and 50 $k$-points in the full 
two-dimensional Brillouin zone (BZ). The muffin-tin radius of 2.5~bohr was used. The band structure of the 
system with SOC is presented in Fig.~2(f). The Wannier functions technique was used on top of self-consistent
DFT calculations to derive an accurate tight-binding Hamiltonian of the system.\cite{Souza:2002, Freimuth:2008, wannier90}
Moreover, we have constructed the matrix elements $\Braket{ \psi_{m\mathbf{k}}| \sigma_\alpha | \psi_{n\mathbf{k}}}$ 
of the Pauli matrices $\sigma_\alpha (\alpha=x,y,z)$, where $\psi_{n\mathbf{k}}$ are the occupied Bloch 
wave functions. These matrices were evaluated from the DFT calculations and transformed into the real space 
representation in terms of maximally localized Wannier functions (MLWFs).\cite{wannier90} This allowed us to 
take into account an exchange field applied perpendicularly to the surface of the bilayer by adding a 
$\sigma_z\cdot B$ term on top of the original Hamiltonian, and to provide an ability to accurately evaluate 
spin polarization $P_\alpha$ (see caption of Fig.~4 for details). In general, applying an exchange field lifts the 
spin degeneracy, where spin-up (spin-down) states get higher (lower) in energy spectra, and the difference of 
the occupation of spin-up and spin down leads to the spin-polarization $P_\alpha$ which can be positive (negative) 
along the direction of exchange fields. In Fig.~2(a)--(e), Fig.~2(k)--(o), and Fig.~4 shown below, spin-up (minority) and 
spin-down (majority) states are indicated in red and blue, respectively.

For a band insulator the transverse anomalous Hall conductivity (AHC) is  $\frac{e^2}{h}$ times the 
conventional integer (first) Chern number, where \cite{Haldane:2004}
\begin{equation}
\mathcal{C}\ =\ \mathcal{C}_{+}\ +\ \mathcal{C}_-\ =\ -\frac{1}{2\pi}\sum_{n=1}^{occ}\int d^2k\,
\Omega_n^{xy}(\mathbf{k}),
\end{equation}
in which the summation goes over all occupied states and $\Omega_n(\mathbf{k})$ is the Berry 
curvature of the $n$-th band, given by
\begin{equation}
\begin{split}
\Omega_n^{xy}(\mathbf{k})&=-2{\rm Im}\Braket{\frac{\partial u_{n\mathbf{k}}}{\partial k_x} | 
\frac{\partial u_{n\mathbf{k}}}{\partial k_y} } \\
   &=-2{\rm Im}\sum_{m\ne n}\frac{\Braket{u_{n\mathbf{k}} | \hat{v}_x | u_{m\mathbf{k}}}
\Braket{u_{m\mathbf{k}}|\hat{v}_y|u_{n\mathbf{k}}}}{(\epsilon_{n\mathbf{k}}-\epsilon_{m\mathbf{k}})^2},
\end{split}
\end{equation}
with $u_{n\mathbf{k}}$ as the lattice-periodic part of $\psi_{n\mathbf{k}}$ with energy eigenvalues 
$\epsilon_{n\mathbf{k}}$, $\hat{v}_\nu$ ($\nu=x,y$) is the velocity operator. To evaluate the spin 
Hall conductivity, we have replaced  $\hat{v}_x$ with the spin-velocity operator 
$\hat{s}_\nu^z=\{\hat{v}_\nu, \sigma_z\}$  in Eq.~(3).

To calculate the spin Chern number, we followed the procedure of Ref.~[\onlinecite{Prodan:2011}].
First, the $\sigma_z$ matrix for the occupied states $\Braket{ \psi_{m\mathbf{k}}| \sigma_z | \psi_{n\mathbf{k}}}$
($m$, $n$ are the band indices of occupied states at each $\mathbf{k}$) is constructed and diagonalized.
Note that due to the spin-flip part of the spin-orbit coupling \cite{Zhang:2011}, the eigenvalues of 
$\sigma_z$ matrix are not necessarily $\pm 1$. Then, using the "spin-up" or "spin-down" eigenvectors 
$|\phi^{\pm}\rangle$ of the $\sigma_z$ matrix, the occupied states are projected into the "spin-up" and 
"spin-down" manifolds $\psi^{\pm}=\Braket{ \phi^{\pm}(\mathbf{k}) | \psi(\mathbf{k})}$, where 
$|\psi(\mathbf{k})\rangle$ denotes the eigenvector of the occupied states. At last, the spin Chern number 
is evaluated by integration of the Berry curvature for each manifold where the derivatives
of the wave functions $\psi^{\pm}$ with respect to $\mathbf{k}$ are obtained using the finite 
difference methods.\cite{Souza:2001} For the spin Chern number, it is only well defined if the 
spectrum of $\sigma_z$ is gapped.\cite{Prodan:2011} In our calculations, we observed that in the 
range of exchange field considered ($ 0\ \text{eV} \le B \le 1\ \text{eV}$) the spectrum of $\sigma_z$ remains 
finite which justifies the usage of spin Chern numbers in this work.

\section{Chern numbers and the phase diagram}

\begin{figure}
\includegraphics[width=8.7cm]{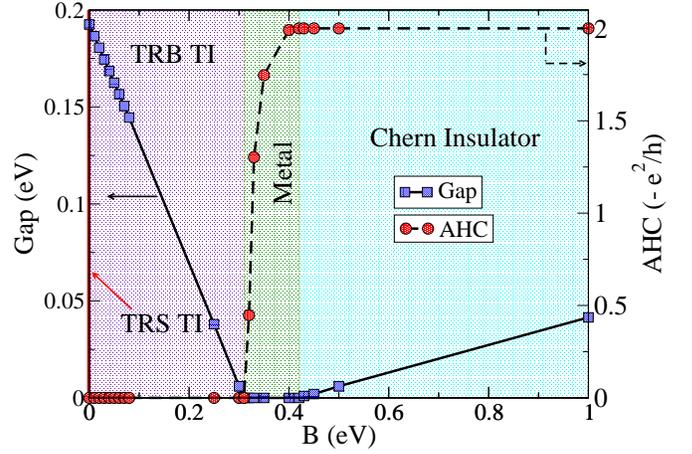}
\caption{
Phase diagram of Bi(111) bilayer with respect to the exchange field $B$. Blue 
squares (red circles) mark the band gap (anomalous Hall conductivity AHC) as a function 
of $B$. Regions of different topological phases, namely, $\mathcal{T}$-conserved TI (TRS TI),
$\mathcal{T}$-broken TI (TRB TI), metal, and Chern insulator (QAH) are shaded differently and 
labeled accordingly.
}
\end{figure}

\begin{figure*}
\includegraphics[width=17.5cm]{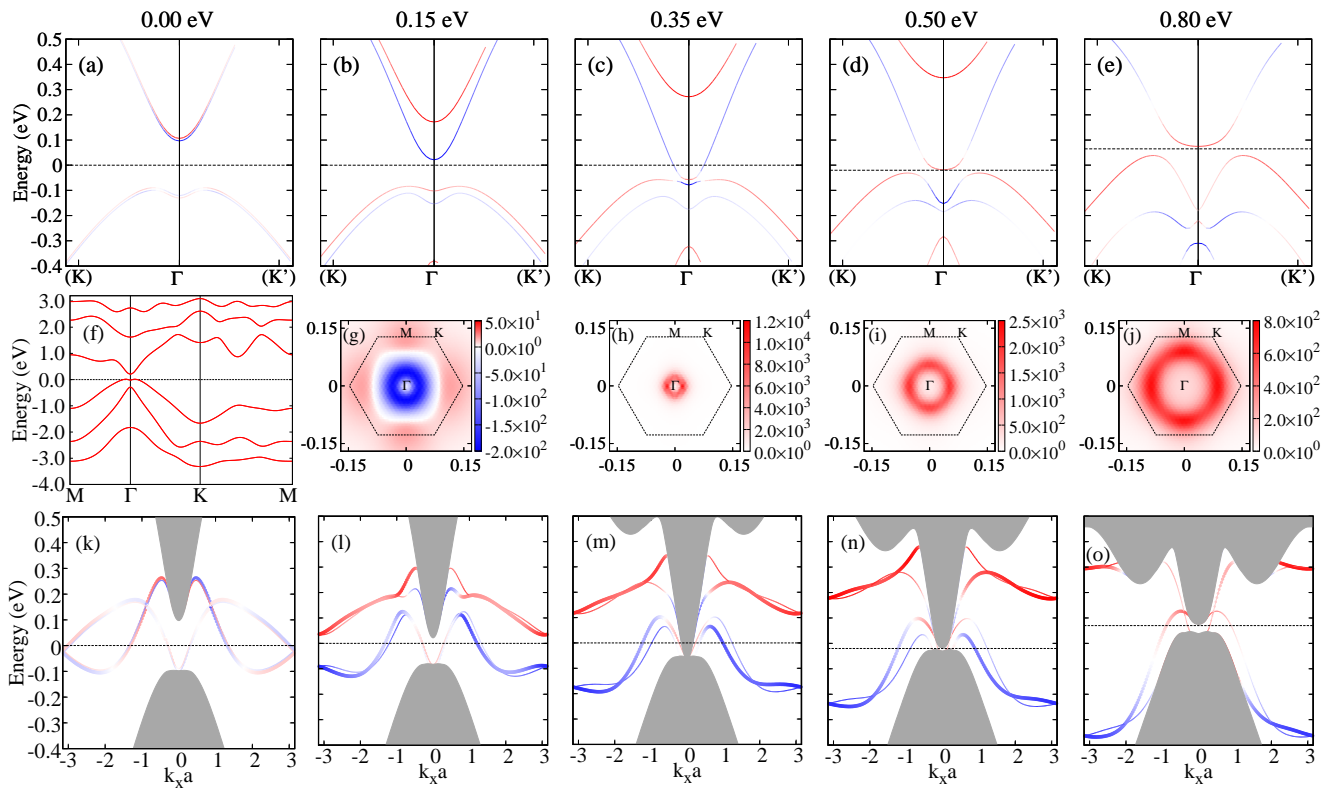}
\caption{
Electronic structure of Bi(111) bilayer in an external exchange field perpendicular to the
surface. (a)--(e) display the 2D "bulk" band structure for an exchange field of 0.0, 0.15, 
0.35, 0.5, and 0.8 eV  (indicated at the top), respectively. Red (blue) color stands for positive (negative)
spin polarization along the field of the states in arbitrary units. The band structures are plotted along the 
K--$\Gamma$--K$^\prime$ path in the vicinity of the $\Gamma$ point.
In (a) the spin-up states are shifted by 10~meV upwards in energy to make the 
degeneracy of the states more visible. In (f) the 2D "bulk" band structure of Bi(111) bilayer is displayed. Each
band is doubly degenerate.
(g)--(j) display the distribution of the Berry curvature of occupied states in the BZ (to a minus sign) with exchange
field marked on the top. Blue (red) color indicates positive (negative) values of the Berry curvature (in a.u.$^{2}$).
The evolution of the band structure of an 80-atom wide zigzag ribbon  is shown in
(k)--(o).  Edge states are colored with the expectation value of the $\sigma_z$, rendered in blue 
and red for negative or positive values, respectively. Size of points indicates the weight from atoms located
on the upper edge of the ribbon.
The gray shaded regions denote the projected 2D "bulk" band structure from 2D BZ onto the 1D $k$ vector 
of the ribbon. In (k) the spin-up states are shifted by 10~meV upwards in energy to make the 
degeneracy of the states more visible. Horizontal dotted lines in (a)--(e) and (k)--(o) indicate the constant energies (chosen to be in 
the bulk gap in insulating phases) for which the spin 
polarization is analyzed as shown in Fig.~4.}
\end{figure*}

Here, we calculate the Chern numbers and analyze the topological phases of Bi(111) bilayer as a function of the 
externally added exchange field $B$, presenting the phase diagram, calculated band gap and AHC in Fig. 1.
When $B=0$, the system is a 2D TI with $\mathbb{Z}_2$ number of $1$ (nontrivial). 
The Chern numbers for the spin-up (minority) and spin-down (majority) manifolds are $\mathcal{C}_\pm =\mp 1$, 
respectively, leading to a spin Chern number
$\mathcal{C}_s=-1$ (cf.~Eq. (1)) in agreement with previous calculations.\cite{Murakami:2006}
In this case, the spin Chern number is equivalent to the $\mathbb{Z}_2$ number. 
Introducing and increasing the exchange field breaks the $\mathcal{T}$-symmetry and causes an exchange splitting 
between the spin-up and spin-down valence and conduction bands, bringing thus the minority valence bands towards 
majority conduction bands,\cite{footnote1} Fig.~2(b).
Further increase of the exchange field leads to a closure of the gap at $B=0.31$ eV, Fig.~2(c). 
As evident from Fig.~2(a)-(c), until the bulk gap is closed, the original inverted band 
structure at the $\Gamma$-point remains topologically nontrivial.
This is confirmed by calculating the spin Chern number $\mathcal{C}_\pm\ =\ \mp 1$ for $0\le B\le 0.31$ eV,
identifying the topological phase of Bi(111) bilayer in this range of the exchange field as the $\mathcal{T}$-broken, or,
TRB TI phase (c.f.~Fig.~1), analogous to that considered in  graphene.\cite{Yang:2011}

\begin{figure}
\includegraphics[width=8.7cm]{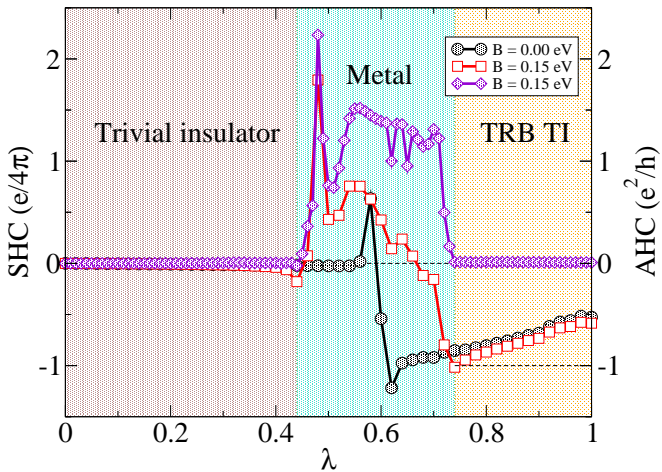}
\caption{
Spin Hall conductivity (SHC) and anomalous Hall conductivity (AHC) of Bi(111) bilayer with 
respect to the scaled strength of the spin-orbit coupling of Bi atoms $\lambda$. $\lambda=1$ corresponds
to the unscaled SOC strength. Black circles (red squares) denote 
the SHC with $B = 0.0$ eV ($B=0.15$ eV), while purple diamonds mark the AHC with $B=0.15$ eV.
Horizontal dashed line at SHC=-$\frac{e}{4\pi}$ stand for the quantized magnitude of SHC if the spin-flip 
part of SOC is switched off.
}
\end{figure}

To verify that the $\mathcal{T}$-broken phase could display non-trivial topological transport properties,
we calculate the transverse
spin Hall conductivity (SHC) as a function of the spin-orbit strength in the system, $\lambda$, and present 
the results in Fig.~3. Consider first the bilayer at $B=0$. At small SOC strength the system is in a trivial
insulator phase, which can be characterized with a zero SHC. Upon increasing $\lambda$, the bilayer 
first goes through a metallic phase, and at around $\lambda=0.7$ of the SOC strength of Bi atoms a transition
to the TI phase with SOC-driven band inversion occurs, which is accompanied by a non-zero SHC of the
magnitude of around $-0.7\frac{e}{4\pi}$.\cite{Murakami:2006} The deviation from the quantized value of 
$-\frac{e}{4\pi}$ is due to the non-vanishing spin-non-conserving part of the spin-orbit interaction: artificially 
switching it off in our calculations\cite{Zhang:2011} we indeed acquire a quantized value of $-\frac{e}{4\pi}$ for the SHC (dashed line in Fig.~3), in 
accordance to the original scenario proposed first for HgTe/(Hg,Cd)Te quantum spin Hall heterostructures.
\cite{Murakami:2004,Bernevig:2006, Konig:2007}. By looking at the SHC 
for the Bi bilayer at $B=0.15$~eV, we observe an almost identical behavior to that without the field, with only 
modifications in the boundaries of the metallic region, SHC in the metallic regime, and a slight difference in 
the value of the SHC at large $\lambda$. Again, by neglecting the spin-non-conserving part of SOC, we 
restore a quantized value of $-\frac{e}{4\pi}$ for the SHC in the TRB TI phase, 
which clearly manifests a topologically non-trivial phase, as far as the topological transport is concerned.
Note that for both phases, TI and TRB TI, the transverse charge conductivity in exactly 
zero when $\lambda$ lies outside of the metallic region. 
The ability to characterize the $\mathcal{T}$-broken TI phase with observable in an experiment non- zero 
spin Hall conductivity could be particularly important for distinguishing this phase from a trivial insulator,
since, as we shall see in the next section, the topological 
protection of the edge states in the system for $B\ne 0$ is lifted.
 
Returning now to the phase diagram in Fig.~1, we observe that the size of the band gap decreases linearly as 
the exchange field is increased, and for $B$ in between 0.31 and 0.42~eV the bulk gap is closed with Bi(111) 
bilayer in the metallic phase, see Fig.~1. However, increasing the exchange field further reopens the gap, which 
results in the occurrence of a topologically non-trivial phase. For $B \ge 0.42$ eV the spin Chern number
of the occupied majority states reverses sign, leading to a zero spin Chern number $\mathcal{C}_s$, 
while the Chern number $\mathcal{C}$ acquires a value of $-$2. That is, in this range of the exchange field
the system is in the QAH, or, Chern insulator phase. It is worth to recall here that in case of a metal in which the 
spin-non-conserving part of the spin-orbit interaction is hypothetically switched off,\cite{Zhang:2011} the intrinsic anomalous Hall 
conductivity is given by the sum of the Hall conductivities for spin-up and spin-down bands, while the spin Hall 
conductivity is given by their difference (spin is a good quantum number in such a case).\cite{Sheng:2005,Zhang:2011}
Evidently, such a correspondence between the Chern number and the spin Chern number,~i.e.,~their representation
as the sum and the difference of the Chern numbers for spin up and spin down bands, holds true for Bi(111) bilayer 
with gapped spectrum of $\sigma_z$, based on the argument that a unitary transformation (eigenvectors of $\sigma_z$ 
in our case) of the occupied states will keep the Chern number invariant. \cite{Resta:1994}

Examining the electronic structure of the system, presented in Fig.~2, reveals that it is the spin-mixing accompanied 
by the exchange of
the Chern number between valence and conduction bands that leads to the phase transition from the
$\mathcal{T}$-conserved TI phase to the QAH phase through the $\mathcal{T}$-broken TI and metallic
phases.  The occurrence of the metallic phase has not been predicted previously~e.g.~for graphene\cite{Yang:2011} 
or Bi$_2$Se$_3$.\cite{Jin:2011} It emerges due  to an overlap in energy of bands with opposite spin character 
in the vicinity of  the $\Gamma$-point (cf. Fig.~2(c)), which is characteristic of a material
with a non-monotonous dispersion of the bands on both sides of the $\Gamma$-point. Such a metallic
phase cannot be achieved~e.g.~for pure Dirac bands, which have a local maximum (minimum) at the high symmetry
(in this case) $\Gamma$-point, as is evident from Fig.~2(c).

Our transport calculations according to Eq.~(3) show that a finite AHC develops in the metallic and QAH phases, 
see Fig.~2. 
Microscopically, the non-vanishing AHC can be traced back to the development of the singular Berry curvature in $k$ 
space as the $B$-field is increased, see Fig.~2(g)-(j). 
For $B$ below 0.31~eV the Berry curvature has a non-trivial distribution in the BZ with the regions of positive and negative
values, comparatively small in magnitude (Fig.~2(g)), while the BZ integral of the Berry 
curvature amounts to zero. 

At the onset of the metallic phase there is a noticeable hybridization between 
the conduction and valence bands of opposite spin, which gives rise to a very large negative contribution to the 
Berry curvature around the points in the BZ at the Fermi level ($E_F$) where the orbital character and the spin character of the 
valence and conduction bands is changing, or, in terminology of Ref.~[\onlinecite{Zhang:2011}], along a "hot loop" 
in the BZ, Fig.~2(c)-(e) and (h)-(j). 
In such a situation, the Berry curvature and the AHC, proportional to the integrated value of the Berry
curvature over the BZ, are dominated by the spin-flip transitions mediated by the non-spin-conserving part
of the spin-orbit interaction.\cite{Zhang:2011} At the $\Gamma$-point, the sign of the Berry curvature 
reverses for $B>0.3$~eV, accompanying the switch in parity of valence and conduction bands.  

In the metallic phase, the AHC as a function of $B$
is monotonously increasing (Fig.~1), while the radius of the hot loop is increasing owing to the shift of the 
hybridization point between the valence and conductance bands away from $\Gamma$ (Fig.~2).
When the gap is reopened for $B > 0.42$ eV, the AHC reaches a value of $-2\frac{e^2}{h}$, with 
the Chern number $\mathcal{C}=-2$, see Fig.~2.
In the QAH phase, we observe from our calculations that out of six valence
bands, the topmost two bands dominate the contribution to the AHC for exchange fields up to 0.60 eV.
When the exchange field is increased further, deeper spin-up bands will be pushed closer to the Fermi
energy and take the spin-down character derived from the conduction bands (cf. Fig.~1(e)), also providing
contribution to the Berry curvature. 

\section{Edge states}

We turn now to the properties of the edge states in Bi(111) bilayer zig-zag ribbon with the
width of 80 atoms in $y$ direction and periodic (infinite) in $x$ direction.\cite{Wada:2011}  The tight-binding Hamiltonian of 
the ribbon was constructed in terms of
the MLWFs by a mere termination of the 2D (infinite) Hamiltonian after 40 unit cells in the $y$ direction.  
For Bi(111) ribbon, when $\mathcal{T}$ symmetry is not
broken, there are twelve edge states in total, as evident from Fig.~2(k).  Six of them are
located on each side of the ribbon, while out of these six there are three "spin-up" left-movers
and three "spin-down" right-movers.  The axis of spin polarization of the edge states is
determined by the details of the electronic structure and spin-orbit interaction, and can be
also energy-dependent.~\cite{Murakami:2006}

\begin{figure}[t!]
\includegraphics[width=8.5cm]{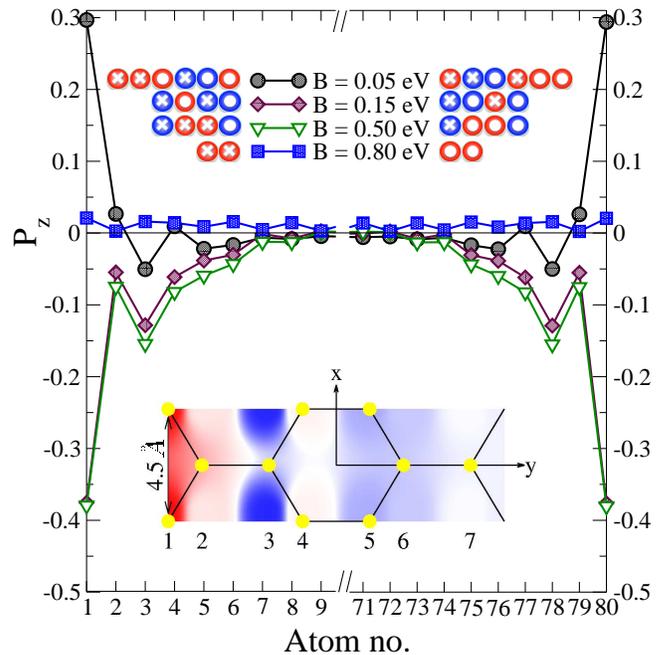}
\caption{ (color online.)
Spatial distribution of the spin polarization $P_z$ at the Fermi energy of the edge states in a zigzag Bi(111) ribbon with 
the width of 80 atoms. $P_z$ is obtained as a sum of the expectation values of $\sigma_z$ for all edge states
at the constant energies (marked with dashed lines in Fig.~2(a)-(e)), multiplied by the weight of the corresponding wavefunction 
on given atom. The structure of the edge states on each side of the ribbon is marked with circles to the left (left edge) 
and to the right (right edge) of the value of the corresponding exchange field $B$. While the direction of propagation
of the edge states is marked with white dots (white crosses) for the direction towards the reader (away from the reader),
the red (blue) color of the circles marks predominantly spin-up (spin-down) character of the edge states.
Lower inset depicts the distribution of $P_z$ for $B=0.05$~eV (in arbitrary units with red for positive and blue for negative
values) along and across the ribbon, where yellow
spheres stand for the positions of Bi atoms.
}
\end{figure}

Applying the exchange field to the ribbon breaks the $\mathcal{T}$-symmetry, lifts the spin
degeneracy and polarizes the edge states along the direction of the applied field, see
Fig.~2(l)-(o). A close inspection of the edge bands in the case of the TRB TI phase, Fig.~2(l) 
reveals that the edge states do not connect the valence and conduction bands in the system, as 
in the case of the edge states of a conventional TI,~c.f.~Fig.~2(k). This means that the Hamiltonian 
of the system can be transformed such that the spectrum at the edges exhibits a gap at the Fermi 
energy. Moreover, owing to the breaking of the $\mathcal{T}$-symmetry, back-scattering is in 
general allowed which results in the dissipative transport properties of the edge states.\cite{Yang:2011} 
Noticeably, in the QAHE phase, Fig.~2(n,o) the connection of the valence and conduction bands of the 
bilayer via the edge states is restored again, resulting in a dissipationless longitudinal transport
of the ribbon, realized by the topologically protected edge states.

Of particular interest here is the spatial distribution of the spin-polarization of the edge states
around the Fermi energy, $P_z$, which plays a very important role in spin-polarized scanning-tunneling microscopy 
(SP-STM) experiments.\cite{Heinze:2006} We calculate this spin polarization as a function of the $B$-field 
in real space across the ribbon, and present the results in Fig.~4. Let us consider first the case of a very 
small exchange field $B=0.05$~eV and look at the spin-polarization at $E_F$ across the ribbon (Fig.~4). 
We observe that the spin polarization is: (i) rather localized at the edges of the ribbon, (ii) reaches very 
large values, (iii) displays oscillations in sign across the ribbon.  These properties can be understood by 
referring to the electronic structure of the edge states for this value of the field, which is quite similar to 
that depicted in Fig.~2(l), only with the smaller separation between the upper (predominantly spin-up) and 
lower (predominantly spin-down) subgroups of two edge states, so that there are twelve states at $E_F$ all 
together. The direction and spin character of the edge states at $E_F$ is sketched in Fig.~4. We observe that 
owing to the presence of the non-zero exchange field and breaking of $\mathcal{T}$ symmetry, the edge-states 
get spin-polarized, while on each side of the ribbon the number of right and left movers is equal for $B < 0.42$. 
This results in zero transverse charge conductance, in accordance to the phase diagram, Fig.~1.

For larger wavevectors  $|k_x\cdot a|\approx\pi$, see Fig.~2(l), the upper group of two states has a strong spin-up
character, and for $B=0.05$~eV (not shown) it crosses the Fermi energy, giving rise to a strong spin-polarization 
in the direction
of the $B$-field. Since the states with larger $k_x$ (away from the bulk bands) are more localized at the edges
of the ribbon than the states with smaller $k_x$ (closer to the bulk bands),\cite{Wada:2011} positive $P_z$  
in Fig.~4 for this field is localized at the ribbon edges. Correspondingly, the states at $E_F$ close to  $\Gamma$
are the source of negative spin polarization, which is smeared more across the ribbon owing to stronger hybridization
with the bulk states, thus giving rise
to the change in the sign of $P_z$ away from the ribbon edges. We speculate that such oscillatory spatial 
dependence of $P_z$ should be a common phenomenon in a situation where the electronic structure of the
edge states of a 2D TI exposed to an exchange field is non-trivial. The spatial distribution of the $P_z$ in
the unit cell for this case is illustrated in the inset of Fig.~4. 

Increasing the exchange field further pushes the tail of long wave length spin-up edge states out of the 
bulk gap completely around $B=0.15$ eV, Fig.~2(l). In this case, there are only 8 edge states
left, whose direction of the spin-polarization at the Fermi energy is opposite to the direction of the $B$-field. 
Since the Bi(111) bilayer exhibits a $\mathcal{T}$-broken TI phase for this value of $B$, there are two left and
two right movers on each side of the ribbon, with zero transported charge, Fig.~4. In correspondence
to the predominantly spin-down character of the edge states in $k$ space, the $P_z$ is large and negative
in the vicinity of the edge of the ribbon. Noticeably, the spin polarization decays much slower with the distance towards the center
of the ribbon, when compared to the case considered previously, owing to the proximity of the edge states in $k$ space 
to the bulk states.
We remark here that the predicted change in sign of $P_z$ is an interesting phenomenon, which can be exploited 
experimentally~e.g.~via tuning the strength of the exchange field by deposition of different adatoms\cite{Jin:2011, Zhang:2012} (see also comments in section V).  

The evolution of the edge states for $0.15\le B\le 0.80$ eV can be seen in Fig.~2(l)-(o). As the exchange field is increased,
the  magnitude of $P_z$ gradually decreases, since the conduction and valence bands become very close in energy, and
the $k_F$ vectors of the edge states  approach the $\Gamma$ point. This is particularly clear for the edge states in the QAH 
phase at $B=0.8$~eV, for which $P_z$ and its decay rate into the ribbon are very small, Fig.~4. For the regime of $B$ 
in between 0.42 and 0.7~eV,~i.e.,~at the onset of the QAH phase, the number of edge states is the same as that in the 
$\mathcal{T}$-broken TI phase at $B=0.15$~eV, but the emergent quantized charge conductance of the edge states 
proportional to $\mathcal{C}$ on each side of the ribbon is evident, Fig.~4. 
When $B > 0.7$ eV four more edge states are pushed out of the bulk gap window, and only four edge states crossing 
$E_F$ are left overall. In this case, we have a situation of two right/left movers on each side of the ribbon, with a small positive
spin polarization, Fig.~4.  The microscopic mechanism for spin-polarized electron conduction in these edge states without possibility to 
back-scatter remains and open and interesting topic for future studies.

\section{Summary}
In this work, based on first principles calculations,  we have considered the emerging topological phases of 
Bi(111) bilayer in an external exchange field $B$. We identify four different phases as the exchange field is varied: 
TI phase for $B=0$, $\mathcal{T}$-broken TI phase for small fields, metallic phase for intermediate fields,
and Chern insulator, or, QAH phase for large exchange fields. We consistently identify each of the insulating phases
in terms of the Chern number $\mathcal{C}$, spin Chern number $\mathcal{C}_s$, spin and charge transverse
conductivity. We attribute the origin of  phase transitions to the spin-orbit mediated spin-mixing between valence 
and conduction bands whose topology in $k$ space is controlled by the applied field.

In particular, we focus on the electronic structure and development  of the edge states in 
a zigzag ribbon of Bi(111) bilayer as a function of the $B$-field. We show that the spin polarization of the edge states along the direction
of the exchange field at the Fermi level can be significant in magnitude, and its sign, as well as its spatial distribution
can be controlled by changing the magnitude of the applied field. Since reversing the direction of the exchange field
leads to reversed propagation direction and reversed spin polarization of the edge states, the exchange field provides
a tool to tune the properties of one-dimensional spin-polarized transport arising at the edges of insulators with non-trivial
topological properties. We also speculate that the spin polarization of the edge states, which characterizes different topological phases,
can be observed in~e.g.~scanning tunneling microscopy experiments.

We remark that the experimental realization of the tunable exchange field applied to a thin Bi(111) bilayer can be sought in
two directions. One way lies in finding a suitable insulating substrate for deposition of Bi(111) which exhibits essential magnetism 
at least at the interface with the bilayer. In this case the direction of the $B$-field can be easily (depending on the magnetocrystalline
anisotropy energy) reversed, while the magnitude of the exchange field can be tuned via~e.g.~the magnetoelectic effect if the
substrate used for deposition has multiferroic properties.  Second route lies in deposition of transition metal adatoms on the
Bi(111) surface. 
Recently, it has been demonstrated theoretically that such adatom deposition can induce considerable exchange splitting with magnitude as large as several tenth of eV on the topological insulator underneath,\cite{Zhang:2012}  and for Bi$_2$Se$_3$, the exchange field is large enough to get the system into the QAH phase.\cite{Jin:2011}.
The magnitude of the exchange field in this case can be controlled by
an appropriate choice of the transition metal, or by deposition of the system on an insulating non-magnetic substrate, which can cause
essential modifications in the distance between the TI and deposited adatoms. We also note, that, generally speaking, the
magnitude of the exchange field necessary to achieve the QAH phase in the phase diagram of Fig.~1 depends directly
on the initial value of the gap in the TI phase without the field.

\section{Acknowledgments}

We acknowledge very helpful discussions with Emil Prodan, Sheng Li, Klaus Koepernik and
Marjana Le\v{z}ai\'c. This work was supported by the HGF-YIG Programme VH-NG-513 and 
by the DFG through Research Unit 912 and grant HE3292/7-1. Computational 
resources were provided by the J\"ulich Supercomputing Centre.


\begin{thebibliography}{99}

\bibitem{Hasan:2010} M.Z. Hasan, and C.L. Kane, Rev. Mod. Phys. {\bf 82}, 3045--3067 (2010).
\bibitem{Qi:2011} X.-L. Qi, and S.-C. Zhang, Rev. Mod. Phys. {\bf 83}, 1057--1110 (2011).
\bibitem{Neto:2009} A.H.Castro Neto, F. Guinea, N.M.R. Peres, K.S. Novoselov, and A.K. Geim, Rev. Mod. Phys. {\bf 81}, 109--162 (2009).
\bibitem{Qi:2008} X.-L. Qi, T.L. Hughes, and S.-C. Zhang, Phys. Rev. B {\bf 78}, 195424 (2008).
\bibitem{Qi:2009} X.-L. Qi, R. Li, J. Zang, and S.-C. Zhang, Science {\bf 323}, 1184--1187 (2009).
\bibitem{Essin:2009} A.M. Essin, J.E. Moore, and D. Vanderbilt, Phys. Rev. Lett. {\bf 102}, 146805 (2009).
\bibitem{Fu:2008} L. Fu, and C.L. Kane, Phys. Rev. Lett. {\bf 100}, 096407 (2008).
\bibitem{Fu:2007} L. Fu, and C.L. Kane, Phys. Rev. B {\bf 76}, 045302 (2007).
\bibitem{Yang:2011} Y. Yang, Z. Xu, L. Sheng, B. Wang, D.Y. Xing, and D.N. Sheng, Phys. Rev. Lett. {\bf 107}, 066602 (2011).
\bibitem{Prodan:2011} E. Prodan, Phys. Rev. B {\bf 83}, 195119 (2011).
\bibitem{Prodan:2010} E. Prodan, New. J. Phys. {\bf 12}, 065003 (2010).
\bibitem{Liu:2008} C.-X. Liu, X.-L. Qi, X. Dai, Z. Fang, and S.-C. Zhang, Phys. Rev. Lett. {\bf 101}, 146802 (2008).
\bibitem{Yu:2010}  R. Yu, W. Zhang, H.-J. Zhang, S.-C. Zhang, X. Dai, and Z. Fang, Science {\bf 329}, 61--64 (2010). 
\bibitem{Chang:2011} C.-Z. Chang, J.-S. Zhang, M.-H. Liu, Z.-C. Zhang, X. Feng, K. Li, L.-L. Wang, X. Chen, X. Dai, Z. Fang, X.-L. Qi, S.-C. Zhang, Y. Wang, K. He, X.-C. Ma, Q.-K. Xue, arxiv:1108.4754.
\bibitem{Jin:2011} H. Jin, J. Im, and A.J. Freeman, Phys. Rev. B {\bf 84}, 134408 (2011).
\bibitem{Haldane:1988} F.D.M. Haldane, Phys. Rev. Lett. {\bf 61}, 2015--2018 (1988).
\bibitem{Novoselov:2005} K.S. Novoselov, A.K. Geim, S.V. Morozov, D. Jiang, M.I. Katsnelson, I.V. Grigorieva, G.V. Dubonos, and A.A. Firsov, Nature {\bf 438}, 197--200 (2005).
\bibitem{Xia:2009} Y. Xia, D. Qian, D. Hsieh, L. Wray, A. Pal, H. Lin, A. Bansil, D. Grauer, Y.S. Hor, R.J. Cava, and M.Z. Hasan, Nature Physics {\bf 5}, 398--402 (2009).
\bibitem{Nagao:2004} T. Nagao, J.T. Sadowski, M.Saito, {\it et al.}, Phys. Rev. Lett. {\bf 93}, 105501 (2004).
\bibitem{Hirahara:2011} T. Hirahara, G. Bihlmayer, Y.Sakamoto, {\it et al.}, Phys. Rev. Lett. {\bf 107}, 166801 (2011).
\bibitem{Wada:2011} M. Wada, S. Murakami, F. Freimuth, and G. Bihlmayer, Phys. Rev. B {\bf 83}, 121310 (R) (2011).
\bibitem{Murakami:2006} S. Murakami, Phys. Rev. Lett. {\bf 97}, 236805 (2006).
\bibitem{Xiao:2011} D. Xiao, W. Zhu, Y. Ran, N. Nagaosa, and S. Okamoto, arxiv:org:1106.4296.
\bibitem{Liu:2011} Z. Liu, C.-X. Liu, Y.-S. Wu, W.-H. Duan, F. Liu, and J. Wu, Phys. Rev. Lett. {\bf 107}, 136805 (2011).
\bibitem{Hohenberg:1964} P. Hohenberg, and W. Kohn, Phys. Rev. {\bf 136}, B864--B871 (1964).
\bibitem{Moruzzi:1978} V.L. Moruzzi, J.F. Janak, A.R. Williams, {\it Calculated Properties of Metals}, New York: Pergamon, (1978).
\bibitem{fleur} For the description of the code see {\tt www.flapw.de}
\bibitem{Souza:2002} X. Wang, J.R. Yates, I. Souza, and D. Vanderbilt, Phys. Rev. B {\bf 74}, 195118 (2006).
\bibitem{Freimuth:2008} F. Freimuth, Y. Mokrousov, D. Wortmann, S. Heinze, and S. Bl\"ugel, Phys. Rev. B {\bf 78}, 035120 (2008).
\bibitem{wannier90} A.A. Mostofi, J.R. Yates, Y.-S. Lee, I. Souza, D. Vanderbilt, and N. Marzari, Comput. Phys. Commun. {\bf 178}, 685 (2008).
\bibitem{Haldane:2004}  F.D.M. Haldane, Phys. Rev. Lett. {\bf 93}, 206602 (2004).
\bibitem{footnote1} This is in contrast to the case considered in
Ref.~[\onlinecite{Jin:2011}], where states of the same spin character from the conduction and valence bands
are pulled to the Fermi energy by increasing the spin-orbit strength.
\bibitem{Souza:2001} I. Souza, N. Marzari, and D. Vanderbilt, Phys. Rev. B {\bf 65}, 035109 (2001).
\bibitem{Murakami:2004} S. Murakami, N.Nagaosa, and S.-C. Zhang, Phys. Rev. Lett. {\bf 93}, 156804 (2004).
\bibitem{Bernevig:2006} B.A. Bernevig, T.L. Hughes, and S.-C. Zhang, Science {\bf 314}, 1757 (2006).
\bibitem{Konig:2007} M.K\"onig, S. Wiedmann, C. Br\"une, {\it et al.}, Science {\bf 318}, 766 (2007).
\bibitem{Sheng:2005} L. Sheng, D.N. Sheng, C.S. Ting, and F.D.M. Haldane, Phys. Rev. Lett. {\bf 95}, 136602 (2005).
\bibitem{Zhang:2011} H. Zhang, F. Freimuth, S. Bl\"ugel, Y. Mokrousov, and I. Souza, Phys. Rev. Lett. {\bf 106}, 117202 (2011).
\bibitem{Resta:1994} R. Resta, Rev. Mod. Phys. {\bf 66}, 899 (1994).
\bibitem{Bruene:2011} C. Bruene, A. Roth, H. Buhmann, E.M. Hankiewicz, L.W. Molenkamp, J. Maciejko, X.-L. Qi, and S.-C. Zhang, arxiv:1107.0585.
\bibitem{Sonin:2011} E.B. Sonin, arxiv:1107.3378
\bibitem{Zhang:2012} H. Zhang, C. Lazo, S. Bl\"ugel, S. Heinze, and Y. Mokrousov, Phys. Rev. Lett. {\bf 108}, 056802 (2012).
\bibitem{Heinze:2006} S. Heinze, Appl. Phys. A {\bf 85}, 407 (2006).

\end{thebibliography}
\end{document}